\documentclass[12pt]{article}

\begin{document}
\baselineskip 6mm

\begin{titlepage}

\hfill\parbox{5cm} {SOGANG-HEP299/02 }

\hfill\parbox{4cm} {{}}

\vspace{25mm}

\begin{center}
{\Large \bf dS/CFT correspondence from the Brick Wall method}

\vspace{15mm}
\textsl{Kyung-Seok Cha\footnote{quantum21@korea.com},
Bum-Hoon Lee\footnote{bhl@ccs.sogang.ac.kr}
and Chanyong Park\footnote{cyong21@hepth.sogang.ac.kr}}
\\[10mm]
{\sl Department of Physics, Sogang University,
Seoul 121-742, Korea} \\
\end{center}

\thispagestyle{empty}

\vskip2cm


\centerline{\bf ABSTRACT} \vskip 4mm \noindent
We research the
entropy of a black hole in curved space-times by 't Hooft`s
approach, so-called the brick wall method. One of these
space-time, a asymptotically dS space-time has two physical horizons; one is a
black hole horizon and the other is a cosmological horizon.
The others have only one horizon, a black hole horizon. Using this
model, we calculate all thermodynamic quantities containing the
background geometric effect and show that entropy is proportional
to area of each boundary. Furthermore, we show that the
Cardy-Verlinde formula can be rederived from the physical
quantities of the brick wall method and this fact becomes a
evidence of the dS/CFT correspondence.

\vspace{2cm}


\end{titlepage}
\pagebreak

\section{Introduction}

Recently, much attention have been focused on studying the de
Sitter (dS) space-time and asymptotically dS space-time. The
motivation of this subject can be described by two aspects. First,
some astronomical observations tell us that our universe with
black hole might be asymptotically de Sitter space-time and
approach to a pure de Sitter space-time as time goes
\cite{Per,CDS,Gar,HKS,FKMP}. Secondly as an analogy of AdS/CFT
correspondence\cite{Malda}, dS/CFT correspondence\cite{S1,no1} between
quantum gravity on de Sitter space-time and the Euclidean
conformal field theory (CFT) on a boundary of the de Sitter
space-time\cite{S2,MS,Hull,HK,Witten,SSV,BHM} was suggested.

Moreover, Verlinde\cite{EK} argued that the Cardy
formula\cite{Cardy}, describing the entropy of a 1+1-dimensional
CFT, can be generalized to an arbitrary dimension and this
generalization gives a so called Cardy-Verlinde formula. This
argument was applied to the quantum gravity in a dS space-time,
which is dual to a certain Euclidean CFT living on the space-like
boundary of the dS space-time and some investigations on this
subject have been carried out
recently\cite{pmi,no2,cno,noo,das,Cai0,Cai1,Cai2,Mann,Halyo,Med,KL,D}.

It is well known that in the dS space-time, there is a
cosmological horizon, which has the similar thermodynamic
properties like the black hole horizon. So the thermodynamics of
the cosmological horizon in asymptotically dS space can be
identified with that of a certain Euclidean CFT living on a
spacelike boundary by the dS/CFT correspondence. But in
cases of the flat and the Anti de Sitter space-time, there is no
such a cosmological horizon so that we can not find such a dS/CFT
correspondence. When we define the Cardy-Verlinde formula, we have
to face a difficulty to calculate some conserved charges including
the mass (total energy) of the gravitational field of the
asymptotically dS space-time. In the spirit of the dS/CFT
correspondence, these conserved charges of the gravitational field
can be identified with those of the corresponding Euclidean CFT.
The difficulty arises due to the absence of the spatial infinity
and the global timelike killing vector. In ref\cite{Cai2}, authors
used the AD mass prescription\cite{AD} on a black hole horizon and
the BBM mass prescription on a cosmological horizon\cite{BBM} so
that the entropy for each horizon can be written as the
Cardy-Verlinde formula. But in this paper, following the brick
wall method\cite{'tHooft,Kim,YLR}, we will calculate the total energy
of the gravitational field. Since the energy obtained here is not
the mass of the black hole but the thermal energy of the gravitational
field near the each horizon, so this is not exactly the same as
the mass of the black hole. Moreover, due to the curved background
geometry the extra term is added to the thermodynamic quantities,
which can be removed in the large radius limit, in other word, the
flat space-time. Although the definition of the total energy is
slightly modified, the similar Cardy-Verlinde formula can be
obtained from this brick wall method. This fact that we can obtain
the similar Cardy-Verlinde formula using various methods, becomes
a good evidence for the dS/CFT correspondence.

The organization of this paper is as follow. In the next section,
we review the brick wall method for Schwarzschild space-time. In
section 3, applying the brick wall method, we calculate
thermodynamic quantities of the curved background space-time and
in the asymptotically dS space-time. In this space-time where
instability occurs, we show that
the asymptotically dS space-time with a black hole
must naurally evolve to a pure dS space-time. Finally, using these
thermodynamic quantities the Cardy-Verlinde formula is obtained. In
section 4, we conclude with some discussion.

\section{Revisited 4-D Schwarzschild space-time}

In this section, we review the brick wall method in
ref\cite{'tHooft}. In ref\cite{'tHooft} 't Hooft contrived the
good exercise to ratiocinate the quantum statistical quantities of
a Schwarzschild black hole. His basic idea is that he identify the
entropy calculated by counting energy states of a particle near
horizon with that of a black hole. However, since the number of
energy level for a particle to occupy at the horizon diverges, for
the energy level to be physically acceptable, one need the UV
cutoff $h_{+}$ of which the role is to make the energy level
finite. Moreover, one must impose an IR cutoff $L$ to make the
infinite volume factor finite in a free energy. This gives us the
consistent result for the Schwarzschild black hole. Somehow, to compare
thermodynamics of 4-D Schwarzschild space-time with the other
space-times with non-zero cosmological constant, we will review
the brick wall method.

Now, we start with the 4-D Schwarzschild black hole metric;
\begin{eqnarray}
ds^2 = -g(r) dt^2 + \frac{1}{g(r)} dr^2 + r^2 d\Omega^2_2,
\end{eqnarray}
where
\begin{eqnarray}
g(r) = 1 - \frac{2M}{r}.
\end{eqnarray}
$M$ is the black hole mass. Since $M>0$, the metric function $g(r)$
has one positive root $R_{+}$ which is the black hole horizon.

To calculate the number of the energy level in the vicinity of the
black hole horizon, we consider the wave equation of massless
scalar field $\Phi$, which describes the quantum fluctuation
of the space-time;
\begin{eqnarray}\label{eq:2.4}
{1\over\sqrt{-g}}\partial_{\mu}(\sqrt{-g}g^{\mu\nu}\partial_{\nu}\Phi)
=0.
\end{eqnarray}
To solve this wave equation, we impose the spherical symmetry
and use the following ansatz
\begin{eqnarray}\label{eq:2.5}
\Phi(x)=e^{-iEt}R(r)Y_{lm}(\theta,\phi) .
\end{eqnarray}
As in ref\cite{'tHooft}, we restrict the wave function to the
region $R_{+}+h_{+}<r<L$;
\begin{eqnarray}\label{eq:2.6}
\qquad\qquad\qquad\qquad\qquad\Phi(x)=0\qquad\qquad\textrm{if}\quad
r \leq R_{+}+h_{+} \quad\textrm{and}\quad r \geq L
\end{eqnarray}
By inserting the equation(\ref{eq:2.5}) into the
equation(\ref{eq:2.4}), we get the radial equation;
\begin{eqnarray}\label{eq:2.7}
\frac{E^2}{g(r)}R(r) +
\frac{1}{r^2}{\partial_{r}}\Big[r^2g(r){\partial_{ r}}R(r) \Big] -
\Big[{l(l+1)\over r^2}\Big]R(r)=0.
\end{eqnarray}

To use WKB approximation, we put $R(r)=A(r)e^{iS(r)}$ where
$A(r)$ is a slowly varying amplitude and $S(r)$ is a fast varying
frequency. Then we look a little complicated equation;
\begin{eqnarray}
\frac{E^2}{g(r)}A(r)&+&g(r)\bigg[\partial^2_{r}A(r)+2i\Big(\partial_{r}A(r)\Big)\Big(\partial_{r}S(r)\Big)
+i\Big(\partial^2_{r}S(r)\Big)A(r)
-\Big(\partial_{r}S(r)\Big)^2A(r)\bigg]\nonumber\\
&+&\frac{2}{r}g(r)\bigg[\partial_{r}A(r)+i\Big(\partial_{r}S(r)\Big)A(r)\bigg]\nonumber\\
&+&\bigg[\partial_{r}g(r)\bigg]\cdot\bigg[\partial_{r}A(r)+i\Big(\partial_{r}S(r)\Big)A(r)\bigg]\nonumber\\
&-&\bigg[\frac{l(l+1)}{r^2}\bigg]A(r)=0.\nonumber
\end{eqnarray}
Finally, if we assume that most contribution of above equation is
the quadratic term of $\partial_{r}S(r)$, we can define the radial
wave number $k^2(r)$ given by $[\partial_rS(r)]^2$;
\begin{eqnarray}\label{eq:2.8}
k^2(r)=\frac{1}{g(r)}\bigg[\frac{E^2}{g(r)}-\frac{l(l+1)}{r^2}\bigg].
\end{eqnarray}
The number of radial mode $n_r$ by counting the number of nodes in
the radial wave function is given by
\begin{eqnarray}\label{eq:2.9}
\pi n_{r}=
\int_{R_{+}+h_{+}}^{L}\sqrt{\frac{r}{(r-2M)}\bigg[\frac{rE^2}{r-2M}-\frac{l(l+1)}{r^2}\bigg]}dr,
\end{eqnarray}
where $L$ and $h_+$ are introduced to regularize the above
integration. If there are no such regulators, some divergences
occur at $r\rightarrow R_{+}(=2M)$ and $r\rightarrow \infty$ in
the equation(\ref{eq:2.9}).

The density of states, $\omega(E_{n})$ at each energy level $E_{n}$
are given by
\begin{eqnarray}\label{eq;g(E)}
\omega(E_{n})&=&\int \pi n_{r}(2l+1)dl\\
&=&\frac{2E^3_{n}}{3}\int_{R_{+}+h_{+}}^{L}\frac{r^2}{g^2(r)}dr.
\nonumber
\end{eqnarray}
In the equation(\ref{eq;g(E)}) the sum over
the angular quantum number $l$ is approximated by integral
and the factor $2l+1$ implies the degeneracy of $L_{z}$.
Hence the free energy,$F$ reads
\begin{eqnarray}\label{eq;2.10}
F &=& \frac{1}{\beta} \sum_{n}\ln(1-e^{-\beta E_{n}}) \nonumber \\
  &\simeq& \frac{1}{\beta} \int_{0}^{\infty}d\omega(E)
         \ln(1-e^{-\beta E})\nonumber\\
  &=&-\frac{2\pi^3}{45\beta^4}\int_{R_{+}+h_{+}}^{L}
         \frac{r^2}{\Big(1-\frac{2M}{r}\Big)^2}dr .
\end{eqnarray}
For convenience, we set
$g(r)=\frac{f(r)}{r}$ and $f(r)=r-2M$.
If we expand $f(r)$ near the horizon( $r=R_{+}+x$ , for small $x>0$ ),
then we get
\begin{eqnarray}\label{eq:2.11}
f(r)&=&f(R_{+})+x f'(R_{+})\nonumber\\
&=&\alpha_{+}x
\end{eqnarray}
where $\alpha_{+}\equiv f'(R_{+})=1$.

Near the horizon, the free energy is rewritten as
\begin{eqnarray}
F&\simeq&-\frac{2\pi^3}{45\beta^4}\int_{h_{+}}^{L-R_{+}}\frac{(R_{+}+x)^4}{\alpha^2_{+}x^2}dx\nonumber\\
&=&-\frac{2\pi^3}{45\beta^4\alpha^2_{+}}\bigg[-\frac{R_{+}}{x}\bigg]_{h_{+}}^{L-R_{+}}
-\frac{2\pi^3}{45\beta^4\alpha^2_{+}}\bigg[\frac{1}{3}x^3\bigg]_{h_{+}}^{L-R_{+}}+\cdots\nonumber
\end{eqnarray}
If we consider only leading terms, we see that the free energy is
\begin{eqnarray}
F&\simeq&-\frac{2\pi^3}{45h_{+}}\bigg(\frac{R^4_{+}}{\alpha^2_{+}\beta^4}\bigg)
-\frac{2\pi^3}{135}\cdot\frac{1}{\alpha^2_{+}\beta^4}\bigg[(L-R_{+})^3-h^3_{+}\bigg].
\end{eqnarray}
Since the first term grows up when $h_{+}$ decreases, it can be
regarded as the contribution of the black hole to the free energy.
The second term contains the finite volume effect. However, we
are only interested in the black hole thermodynamics and so the second
term proportional to the volume is ignored. Then, the free
energy of the black hole, $F_{+}$ is given by
\begin{eqnarray}
F_{+}=-\frac{2\pi^3}{45h_{+}}\bigg(\frac{R^4_{+}}{\alpha^2_{+}\beta^4_{+}}\bigg)
\end{eqnarray}
where the inverse temperature $\beta_{+}$ at horizon is
\begin{eqnarray}
\frac{1}{\beta_{+}}= T_+ &=&\frac{1}{4\pi}\bigg[\frac{d}{dr}g(r)\bigg]_{r=R_{+}}\nonumber\\
&=&\frac{\alpha_{+}}{4\pi R_{+}}.
\end{eqnarray}
Using the usual thermodynamic formula, we find that the total
energy $U_{+}$ and the entropy $S_{+}$ are
\begin{eqnarray}
U_{+}&=&\frac{\partial}{\partial\beta_{+}}(\beta_{+} F_{+})\nonumber\\
&=&\frac{2\pi^3}{15h_{+}}\bigg(\frac{R^4_{+}}{\alpha^2_{+}\beta^4_{+}}\bigg)\\
S_{+}&=&\beta_{+}(U_{+}-F_{+})=\beta^2_{+} \frac{\partial
F_{+}}{\partial \beta_{+}}\nonumber\\
&=&\frac{8\pi^3}{45h_{+}}\bigg(\frac{R^4_{+}}{\alpha^2_{+}\beta^3_{+}}\bigg)
\end{eqnarray}
For the entropy satisfying the area law\cite{Bek,Hawking}, we take
the UV cutoff $h_{+}$ to be $1/360\pi R_{+}$. Accordingly, the
free energy $F_{+}$, the total energy $U_{+}$ and the entropy
$S_{+}$ result into
\begin{eqnarray}
F_{+}&=&\frac{-1}{16}\alpha_{+}R_{+}
\end{eqnarray}
\begin{eqnarray}
U_{+}&=&\frac{3}{16}(\alpha_{+}R_{+})\nonumber\\
&=&\frac{3}{8} M \\
S_{+}&=&\pi R^2_{+}=\frac{A_{+}}{4}
\end{eqnarray}
where $A_{+}$ is the area of the black hole horizon. Notice that
$U$ is the thermal energy of the quantum fluctuations and this is
not equal to the black hole mass.

Next, we calculate the invariant distance of the brick wall
which is independent of
coordinates as ref\cite{'tHooft}.
\begin{eqnarray}
\int_{R_{+}}^{R_{+}+h_{+}}ds
&=&\int_{R_{+}}^{R_{+}+h_{+}}\sqrt{\frac{r}{r-2M}}dr \nonumber\\
&\simeq& \sqrt{\frac{1}{90\pi}}
\end{eqnarray}

\section{Thermodynamics of the black hole}

\subsection{Thermodynamic quantities}

In this section, we will consider more general black hole metric in the curved
background space-time. The metric is given by
\begin{eqnarray}
ds^2 = -g(r) dt^2 + \frac{1}{g(r)} dr^2 + r^2 d\Omega^2_2,
\end{eqnarray}
where
\begin{eqnarray}
g(r) = 1 - \frac{2M}{r}+\frac{Q^2}{r^2} - k \frac{r^2}{l^2}.
\end{eqnarray}
Here, $M$ and $Q$ are the mass and the charge of the black hole
respectively and $l$ is a space-time curvature radius. $k$ implies
the sign of the cosmological constant: $k>0$ and $k<0$ mean a de
Sitter (dS) and a Anti de Sitter (AdS) space-time respectively and
$k=0$ is a asymptotically flat space-time.

Note that if we choose a infinite radius ($l \to \infty$) in the
case $k \ne 0$, the dS and AdS go to the flat space-time ($k=0$).
So if we will have calculated physical quantities of the black
hole in the non-zero cosmological constant background space-time,
in the limit of $l\to \infty$ these quantities must be consistent
with those of the asymptotically flat space-time. Using this fact,
we can indirectly prove that our calculations in curved background
are correct. Though a black hole with a mass $M$ and a charge $Q$
has usually two horizons, the large one of these is called an
event horizon of the black hole, $R_+$ determined by
\begin{equation} \label{eqR+}
g(R_{+})=1-\frac{2M}{R_{+}}+\frac{Q^2}{R^2_{+}}-\frac{k R^2_{+}}{l^2}=0 ,
\end{equation}
which is the radius of the black hole. Since in the inside of the event horizon
there is no time-like curve, we consider the outside of the event horizon, $r>R_+$.
For the dS space-time ($k>0$), there exists another horizon
called a cosmological horizon $R_c$
determined by
\begin{equation} \label{eqRc}
g(R_{c})=1-\frac{2M}{R_{c}}+\frac{Q^2}{R^2_{c}}-\frac{R^2_{c}}{l^2}=0
\end{equation}
where we set $k=1$. Since the time-like curve can be defined
in the inside of the cosmological horizon, for the dS space-time with a black hole,
we consider the time-like region $R_+ < r <R_c$
only.
When $R_+ = R_c = l/\sqrt{3}$ for $Q=0$, this black hole becomes a
Narai black hole. Since unlike a dS space-time a flat and a AdS space-time have
no such a cosmological horizon, the time-like region is given by $r>R_+$.

To probe the energy level of a particle near one or two horizons, we
consider the wave equation of massless scalar field $\Phi$;
\begin{eqnarray}\label{eq:3.4}
{1\over\sqrt{-g}}\partial_{\mu}(\sqrt{-g}g^{\mu\nu}\partial_{\nu}\Phi)=0.
\end{eqnarray}
The solution ansatz of this wave equation under the spherical
symmetry is given by
\begin{eqnarray}\label{eq:3.5}
\Phi(x)=e^{-iEt}R(r)Y_{lm}(\theta,\phi).
\end{eqnarray}
Here we restrict the wave function to the region
$R_{+}+h_{+}<r<R_{c}-h_{c}$;
\begin{eqnarray}\label{eq:3.6}
\qquad\qquad\qquad\qquad\qquad\Phi(x)=0\qquad\textrm{if}\quad r
\leq R_{+}+h_{+} \quad\textrm{and}\quad r \geq R_{c}-h_{c}
\end{eqnarray}
By inserting the equation(\ref{eq:3.5}) into the
equation(\ref{eq:3.4}), we get the radial equation
\begin{eqnarray}\label{eq:3.7}
\frac{E^2}{g(r)}R(r) +
\frac{1}{r^2}{\partial_{r}}\Big[r^2g(r){\partial_{ r}}R(r) \Big] -
\Big[{l(l+1)\over r^2}\Big]R(r)=0.
\end{eqnarray}
Using the similar WKB approximation in section 2, the radial wave
number is given by
\begin{eqnarray}\label{eq:3.8}
k^2(r)=\frac{1}{g(r)}\bigg[\frac{E^2}{g(r)}-\frac{l(l+1)}{r^2}\bigg],
\end{eqnarray}
which depends on the metric.
The equation(\ref{eq:3.8}) means that the frequency of a particle
near the horizon is highly excited and diverge at the horizon.
These infinities are the UV divergences similar to problem in
renormalization of QFT contents\cite{Myers,Solo,BL}. Hence we need
some UV cutoffs: for the dS space-time there are two cutoffs,
$h_{+}$ and $h_{c}$ and for the flat and the AdS space-time there
is only one cutoff. The introduction of the cutoff makes the
number of energy state of $\Phi$ physically meaningful. Since in
the asymptotically dS space-time the infinite volume factor does
not occur, we do not have to introduce the IR cutoff. But in the
asymptotically flat and AdS space-time, there is no cosmological
horizon and so we need an IR cutoff to regularize the infinite
volume effect which is not important. Hence we can interpret $R_c$
as an IR cutoff and in this case, $h_c$ must be set to zero in the
asymptotically AdS and flat space-time. Under the restriction on
the wave function to $R_{+}+h_{+}<r<R_{c}-h_{c}$, the number of
radial mode $n_r$ by counting the number of nodes in the radial
wave function is given by
\begin{eqnarray}
\pi n_{r}= \int_{R_{+}+h_{+}}^{R_{c}-h_{c}}k(r, l, E)dr .
\end{eqnarray}
And at each energy level $E_{n}$, the number of state $\omega(E_{n})$ is given by
\begin{eqnarray}\label{eq;w(E)}
\omega(E_{n})&=& \int \pi n_{r}(2l+1)dl \nonumber \\
&=&\frac{2 E^3_{n}}{3}\int_{R_{+}+h_{+}}^{R_{c}-h_{c}}\frac{r^2}{g^2(r)}dr.\nonumber
\end{eqnarray}
Hence $F$ is
\begin{eqnarray}
F &=&
-\frac{2\pi^3}{45\beta^4}\int_{R_{+}+h_{+}}^{R_{c}-h_{c}}\frac{r^2}{g^2(r)}dr .
\end{eqnarray}

For the dS space-time ($k=1$), the integration part is rewritten as
\begin{equation}\label{eq;4.2}
\int_{R_{+}+h_{+}}^{R_{c}-h_{c}}\frac{r^2}{\Big(1-\frac{2M}{r}+\frac{Q^2}{r^2}
         -\frac{r^2}{l^2}\Big)^2}dr
=\int_{R_{+}+h_{+}}^{L}\frac{r^6}{f^2(r)}dr
+\int_{L}^{R_{c}-h_{c}}\frac{r^6}{f^2(r)}dr
\end{equation}
where for convenience we set $g(r)=\frac{f(r)}{r^2}$ and
$f(r)=\Big(r^2-2Mr+Q^2-\frac{r^4}{l^2}\Big)$. Note that $L$ does not represent the IR
cutoff. In the cases of the flat ($k=0$) and the AdS space-time ($k=-1$),
because there is no cosmological horizon
the integration part is written as
\begin{equation}\label{eq;4.21}
\int_{R_{+}+h_{+}}^{R_{c}}\frac{r^2}{\Big(1-\frac{2M}{r}+\frac{Q^2}{r^2}
         -\frac{k r^2}{l^2}\Big)^2}dr
=\int_{R_{+}+h_{+}}^{L}\frac{r^6}{f^2(r)}dr
+\int_{L}^{R_{c}}\frac{r^6}{f^2(r)}dr .
\end{equation}
Here $R_c$ in these flat and AdS background space-times is not a
physical radius but an IR cutoff and so the second term of the
above equation gives term proportional to the volume at leading
order. Because this volume effect must be removed from our
calculations like the Schwarzschild black hole case,
all thermodynamic quantities come from the first
term of the above equation, in other words, the calculation near
the event horizon of the black hole $R_+$.

Now in the dS space-time,
we expand $f(r)$ near two horizons $R_{+}$ and $R_{c}$.
Near the event horizon $r=R_{+}+x$, the function $f(r)$ is given by
\begin{equation}
f(r) \simeq \alpha_{+} x,
\end{equation}
and near the cosmological horizon $r=R_{c}-x$, $f(r)$ becomes
\begin{equation}
f(r) \simeq -\alpha_c x,
\end{equation}
where $\alpha_{+}\equiv(2R_{+}-2M-\frac{4k}{l^2}R^3_{+})>0$ and
$\alpha_{c}\equiv(2R_{c}-2M-\frac{4}{l^2}R^3_{c})<0$.
Notice that since only the dS space-time has the cosmological horizon,
$\alpha_c$ appears in the case of $k>0$ only and
we can set $k=1$ in the definition of $\alpha_c$.
If $M$ is considered as a function of $R_+$ and $R_c$ these two factors can be
rewritten as
\begin{eqnarray}
\alpha_{+}=\bigg(R_{+}-\frac{Q^2}{R_{+}}-\frac{3k R^3_{+}}{l^2}\bigg)>0\quad\textrm{and}\quad
\alpha_{c}=\bigg(R_{c}-\frac{Q^2}{R_{c}}-\frac{3 R^3_{c}}{l^2}\bigg)<0\nonumber
\end{eqnarray}
Consequently, the
equation(\ref{eq;4.2}) for the asymptotically dS space-time is given by
\begin{eqnarray}
&& \frac{1}{\alpha_{+}^2}\int_{h_{+}}^{L-R_{+}}\frac{(x+R_{+})^6}{x^2}dx+
\frac{1}{\alpha_{c}^2}\int_{R_{c}-L}^{h_{c}}\frac{(R_{c}-x)^6}{x^2}d(-x)\nonumber \\
&& \quad  = \frac{R^6_{+}}{\alpha^2_{+}}\bigg[-\frac{1}{x}\bigg]_{h_{+}}^{L-R_{+}}+
\frac{5R^2_{+}}{\alpha^2_{+}}\bigg[x^3\bigg]_{h_{+}}^{L-R_{+}}+
\frac{6R^5_{+}}{\alpha^2_{+}}\bigg[\log{x}\bigg]_{h_{+}}^{L-R_{+}} \nonumber\\
&& \quad \;\;\; + \frac{R^6_{c}}{\alpha^2_{c}}\bigg[-\frac{1}{x}\bigg]_{h_{c}}^{R_{c}-L}+
\frac{5R^2_{c}}{\alpha^2_{c}}\bigg[x^3\bigg]_{h_{c}}^{R_{c}-L}-
\frac{6R^5_{c}}{\alpha^2_{c}}\bigg[\log{x}\bigg]_{h_{c}}^{R_{c}-L}+\cdots \nonumber
\end{eqnarray}
where the second and the fourth terms are proportional to the volume.
Therefore at the leading order the free energy becomes
\begin{eqnarray}
F &=& F_{+}+F_{c} \nonumber \\
F_{+}&=&-\frac{2\pi^3}{45h_{+}}\bigg(\frac{R^6_{+}}{\alpha^2_{+}\beta^4_{+}}\bigg)
\nonumber \\
F_{c}&=&-\frac{2\pi^3}{45h_{c}}\bigg(\frac{R^6_{c}}{\alpha^2_{c}\beta^4_{c}}\bigg),
\end{eqnarray}
where $F_+$ and $F_c$ are originated from of the black hole and the dS space-time respectively.
In the flat and AdS space-time, due to the absence of the cosmological horizon,
$F_c$ does not appears. Hence the free energy of the black hole in the asymptotically
flat and AdS space-time
is given by $F=F_+$.
Applying the thermodynamic formula to calculate the temperature $T$, the total
energy $U$ , the chemical potential $\varphi$ and the entropy $S$,
we find that near the event horizon of the black hole
\begin{eqnarray}
\frac{1}{\beta_{+}}&\equiv&T_{+} \nonumber \\
&=&\frac{\alpha_{+}}{4\pi R^2_{+}} \\
\nonumber \\
U_{+}&=&\frac{\partial}{\partial\beta_{+}}(\beta_{+}F_{+})\nonumber\\
&=&\frac{2\pi^3}{15h_{+}}\bigg(\frac{R^6_{+}}{\alpha^2_{+}\beta^4_{+}}\bigg) \\
\nonumber\\
\varphi_{+}&=&\bigg(\frac{\partial}{\partial Q}F_{+}\bigg)_{T_{+}} \nonumber\\
&=& -\frac{2Q}{R_{+}}\bigg(\frac{4\pi^3}{45h_{+}}\cdot\frac{R^6_{+}}{\beta^4_{+}
\alpha^3_{+}}\bigg)\\
\nonumber\\
S_{+}&=&\beta^2_{+} \frac{\partial F_{+}}{\partial \beta_{+}}\nonumber\\
&=&\frac{8\pi^3}{45h_{+}}\bigg(\frac{R^6_{+}}{\alpha^2_{+}\beta^3_{+}}\bigg)
\end{eqnarray}
and near the cosmological horizon of the asymptotically dS space-time
\begin{eqnarray}
\frac{1}{\beta_{c}}&\equiv&T_{c} \nonumber \\
&=& \frac{-\alpha_{c}}{4\pi R^2_{c}} \\
\nonumber \\
U_{c}&=&\frac{\partial}{\partial\beta_{c}}(\beta_{c}F_{c})\nonumber\\
&=&\frac{2\pi^3}{15h_{c}}\bigg(\frac{R^6_{c}}{\alpha^2_{c}\beta^4_{c}}\bigg)\\
\nonumber\\
\varphi_{c}&=&\bigg(\frac{\partial}{\partial Q}F_{c}\bigg)_{T_{c}} \nonumber \\
&=& -\frac{2Q}{R_{c}}\bigg(\frac{4\pi^3}{45h_{c}}\cdot\frac{R^6_{c}}{\beta^4_{c}
\alpha^3_{c}}\bigg) \\
\nonumber \\
S_{c}&=&\beta^2_{c} \frac{\partial F_{c}}{\partial \beta_{c}}\nonumber\\
&=&\frac{8\pi^3}{45h_{c}}\bigg(\frac{R^6_{c}}{\alpha^2_{c}\beta^3_{c}}\bigg).
\end{eqnarray}
Notice that all thermodynamic quantities are locally defined at each horizon.
So in the case of the asymptotically dS space-time containing the black hole
there are two temperatures,
which cause the instability of this system. In the next section, we will comment
this instability.

In the same way in section 2, we choose the invariant distances of the brick walls
\begin{eqnarray}
h_{+}=\frac{\alpha_{+}}{360\pi R^2_{+}}  \qquad\textrm{and}\qquad
h_{c}=\frac{-\alpha_{c}}{360\pi R^2_{c}} ,
\end{eqnarray}
then we find that the free energy $F_{+}$ , the total energy
$U_{+}$ , the chemical potential $\varphi_{+}$ and the entropy
$S_{+}$ near the event horizon are
\begin{eqnarray}\label{eq;U+}
F_{+}&=&\frac{-1}{16}\bigg(R_{+}-\frac{Q^2}{R_{+}}-\frac{3R^3_{+}}{l^2}\bigg)\\
U_{+}&=&\frac{3}{16}\bigg(R_{+}-\frac{Q^2}{R_{+}}-\frac{3R^3_{+}}{l^2}\bigg)\\
\varphi_{+}&=&-\frac{Q}{4R_{+}}\\
S_{+}&=&\pi R^2_{+}
\end{eqnarray}
and those near the cosmological horizon are
\begin{eqnarray}\label{eq;Uc}
F_{c}&=&\frac{1}{16}\bigg(R_{c}-\frac{Q^2}{R_{c}}-\frac{3R^3_{c}}{l^2}\bigg)\\
U_{c}&=&-\frac{3}{16}\bigg(R_{c}-\frac{Q^2}{R_{c}}-\frac{3R^3_{c}}{l^2}\bigg)>0\\
\varphi_{c}&=&\frac{Q}{4R_{c}}\\
S_{c}&=&\pi R^2_{c} .
\end{eqnarray}
Notice that the entropy at each horizon satisfies the area law.

To convince that our result is reasonable, take the limit of
the zero cosmological constant limit, $\Lambda \to 0$ (the curvature radius
$l \to \infty$).
In the SdS (Schwarzschild dS) black hole ($Q=0$) case, using the equation
(\ref{eqR+}), the total energy $U_+$ can be rewritten as
\begin{equation} \label{bgeff}
U_+ = \frac{3}{8} \bigg( M - \frac{2R^3_{+}}{l^2} \bigg) .
\end{equation}
In the limit $l \to \infty$, which in the equation (\ref{bgeff})
means that the dS space-time goes to the flat space-time, $U_+$ is
exactly equal to that of Schwarzschild black hole in section 2 and
$R_+ \to 2M$. As mentioned before, in flat space-time
the cosmological horizon $R_{c}$ must be interpreted as the IR
cutoff and so we have to ignore all quantities
$F_{c}$, $U_{c}$ and $S_{c}$ near the cosmological horizon. As a
result, in this limit we can check that the total energy $U_{+}$
and entropy $S_{+}$ are the same as those of the Schwarzschild
black hole \cite{'tHooft};
\begin{eqnarray}\label{eq;3.20}
U_{+}=\frac{3}{8}M \qquad\textrm{and}\qquad S_{+}=\pi(R_+)^2 .
\end{eqnarray}

In ref\cite{Cai1,Cai2}, the black hole mass was identified with
the energy of the black hole and the dS space-time, for example
$U_+ = M$ and $U_c =-M$, which are different with our results. For
the Schwarzschild black hole case, the used metric is valid to
observer who live in the asymptotically flat space-time and so the
black hole mass $M$ is a special quantity observed by such an
observer. In generally curved space-time, since there is no global
timelike killing vector, it is not clear how to
define the black hole mass. But if the black hole mass is defined,
it has to be that of the Schwarzschild black hole in the large $l$ limit
and contain the effect of the background
geometry. However, in equation (\ref{bgeff}), the first term, which
survives in the limit of $l \to \infty$, is the effect of the
black hole in the asymptotically flat space-time and is the
exactly same as the result of \cite{Cai1,Cai2} up to $\frac{3}{8}$
factor. The second term is the background geometric effect which
vanishes in the flat space-time. Therefore, if we ignore the
factor $\frac{3}{8}$ - whose origin is not clear but we guess that
it comes from the thermalization of the quantum fluctuations,
all thermodynamic quantities obtained here are the most reliable.

\subsection{Instability of the black hole in the dS space-time}

As mentioned in the previous section, in the asymptotically dS
space-time containing a black hole there are two temperatures
defined locally at each horizon. These two temperatures imply that
this system is not an equilibrium state and so the energy transfer
can occur. In our calculation, it is shown that at the leading
order the total entropy of this system is given by the sum of
those at each horizon. For convenience considering only the
Schwarzschild dS space-time ($Q=0$), the total entropy at leading
order is given by
\begin{eqnarray}\label{eq;3.23}
S=S_{+}+S_{c}=\pi R^2_{+}+\pi R^2_{c}.
\end{eqnarray}
From the equation(\ref{eqR+}) and the equation(\ref{eqRc}), we get
\begin{eqnarray}\label{eq;3.24}
R^2_{+}+R_{c}R_{+}+R^2_{c}-l^2=0
\end{eqnarray}
Inserting the equation(\ref{eq;3.24}) into (\ref{eq;3.23}) and
after some calculation, $S$ is described as the function of
$R_{c}$;
\begin{eqnarray}\label{eq;3.25}
S&=&\pi l^2-\pi R_{c}R_{+}\nonumber\\
&=&\pi l^2+\frac{\pi}{2}\bigg(R^2_{c}-R_{c}\sqrt{4 l^2-3R^2_{c}}\bigg)
\end{eqnarray}
where $\frac{l}{\sqrt{3}}<R_{c}<l$ and $S$ have no extremum
value in this region. Consequently, when $R_{c}\to l$,
$S$ approaches to maximum. So the maximal entropy
$S_{max}$ reads
\begin{eqnarray}\label{eq;3.26}
S\rightarrow S_{max}=\pi l^2.
\end{eqnarray}
Note that this maximal entropy is the same as that of the pure dS
space-time. According to the thermodynamic law, since all
physical system evolves in the direction of increasing the
entropy, the Schwarzschild dS space-time evolves to the pure dS
space-time. This means that the black hole disappears due to its
evaporation which decreases the mass and makes the cosmological
horizon $l$. This result is equal to that of ref\cite{Halyo}.
Therefore the Schwarzschild dS space-time is unstable, while the
pure dS space-time is stable.

When the mass of the black hole $M$ increases, the event horizon
expands and the the cosmological horizon shrinks. Finally when the
event horizon meets the cosmological horizon $R_{c} =R_+  =
\frac{l}{\sqrt{3}}$, the black hole becomes the Narai black hole
which have the maximal size in the dS space-time. Although the
Narai black hole by itself has the maximal entropy, the total
entropy of this space-time becomes a minimal value;
\begin{eqnarray}\label{eq;3.27}
S_{min} = \frac{2\pi}{3} l^2 .
\end{eqnarray}

\subsection{ The Cardy-Verlinde formula}

In this subsection, the Cardy-Verlinde formula in
ref\cite{EK} is rederived from the brick wall method.
The total energy $U(S,V)$ may be written as a sum of
two terms
\begin{eqnarray}
U(S,V)=E_{ext}(S,V)+E_{sub}(S,V)
\end{eqnarray}
where $E_{sub}(S,V)$ is the sub-extensive energy which can contains
not only the casimir energy $E_c (S,V)$ but also other sub-extensive energy.
In the general charged dS black hole, the total energy near the black hole is given by
\begin{eqnarray}
U=E_{ext}+E_{Q}+\frac{1}{3}E_{C} ,
\end{eqnarray}
where
\begin{eqnarray}
E_{ext}&=&-\frac{9}{16 l^2}R^3_{+}\nonumber\\
E_{Q}&=&-\frac{3Q^2}{16R_{+}}\nonumber\\
\frac{E_{C}}{3}&=&\frac{3}{16}R_{+} .
\end{eqnarray}
Notice that $E_Q$ is another sub-extensive energy and the factor
$\frac{1}{3}$ in front of the casmir energy is included because we
consider four-dimensional space-time with three spatial
dimensions. From these energies, we find the Cardy-verlinde
entropy $(S_{+})_{C-V}$
\begin{equation}
(S_{+})_{C-V}=\frac{2\pi l}{3}\sqrt{\frac{E_{C}}{3}\bigg|\{3(U_{+}-E_{Q})-E_{C}\}\bigg|} .
\end{equation}
After more some calculation, we find that the Cardy-Verlinde
entropy is equal to the entropy obtained in the brick wall method,
${S_+}_{b \cdot w}$, up to some factor;
\begin{equation}
(S_{+})_{C-V} = \frac{3}{8}(S_{+})_{b\cdot w}.
\end{equation}

Similarly we reproduce above the procedure near the cosmological horizon.
From the total energy(\ref{eq;Uc}) near the cosmological horizon,
\begin{equation}
U=E_{ext}+E_{Q}+\frac{1}{3}E_{C}
\end{equation}
where
\begin{eqnarray}
E_{ext}&=&\frac{9}{16 l^2}R^3_{c}\nonumber\\
E_{Q}&=&\frac{3Q^2}{16R_{c}}\nonumber\\
\frac{E_{C}}{3}&=&-\frac{3}{16}R_{c} ,
\end{eqnarray}
the relation between the Cardy-Verlinde entropy $(S_{c})_{C-V}$ and
the brick wall entropy $(S_{c})_{b\cdot w}$ is given by
\begin{eqnarray}
(S_{c})_{C-V}&=&\frac{2\pi l}{3}\sqrt{\bigg|\frac{E_{C}}{3}\bigg|\{3(U_{c}-E_{Q})-E_{C}\}}\nonumber\\
&=&\frac{3}{8}(S_{c})_{b\cdot w} .
\end{eqnarray}
Here using the independent way, that is, the brick wall method, we have obtained the thermodynamical
quantities. Since these quantities, especially the entropy, satisfy the Cardy-Verlinde
formula, the result in this paper can be a good evidence for the dS/CFT correspondence.

\section{Discussion}

Using the 't Hooft's brick wall approach, we have calculated all
thermodynamic quantities of asymptotically flat and non-flat
space-time except rotating space-time. Asymptotically de Sitter
space-time has two physical horizons; one is a event horizon of
the black hole  and the other is a cosmological horizon. And the
others have the only event horizon of the black hole. The
thermodynamic quantities at each horizon is similar to those of
other papers. But since instead of the black hole mass the thermal
energy of the gravitational field near the horizons is defined as a
total energy, the energy is slightly modified containing the
factor $3/8$ which, we guess, is the thermalization factor and
the background geometrical effect. These geometrical effect
disappears in the large radius limit(the flat space-time) and in
this limit, if we ignored the factor $3/8$ which always appears in
the brick wall method, the energy obtained here is exactly the
same as the black hole mass. Generally the mass of the
Schwarzschild black hole is defined by the observer living in the
flat space-time which is the asymptotic space-time of the
Schwarzschild black hole. So in the large radius limit, the
results of the curved space-time is exactly the same as those of
the flat space-time. In this paper, we have shown that our result
is consistent with these fact. Therefore our result is reasonable
and is the generalized one containing the background geometric
effect.

Moreover, in spite of some different results, it is very
interesting that the Cardy-Verlide formula can be rederived from
these results up to some factor. This fact is a good evidence for
the dS/CFT correspondence. \\

\noindent {\bf  Acknowledgments}
This work was supported by grant No. R01-1999-00018 from the
Interdisciplinary Research Program of the KOSEF.
BHL is also supported by the Sogang University Research Grant
in 2001.

\end{document}